# Optimizing Sorting of Micro-Sized Bio-Cells in Symmetric Serpentine Microchannel using Machine Learning

Sayan Karmakar[1], Md Safwan Mondal[2], Anish Pal[2], Sourav Sarkar[3]

[1]Department of Civil Engineering, Jadavpur University, Kolkata-700032, India
[2]Department of Mechanical Engineering, University of Illinois-Chicago, Chicago, U.S.A.
[3]Department of Mechanical Engineering, Jadavpur University, Kolkata-700032, India

**ABSTRACT**

Efficient sorting of target cells is crucial for advancing cellular research in biology and medical diagnostics. Inertial microfluidics, an emerging technology, offers a promising approach for label-free particle sorting with high throughput. This paper presents a comprehensive study employing numerical computational fluid dynamics (CFD) simulations to investigate particle migration and sorting within a symmetric serpentine microchannel. By adopting a Eulerian approach to solve fluid dynamics and a Lagrangian framework to track particles, the research explores the impact of flow Reynolds number and the number of loops in the serpentine channel on sorting efficiency. To generate a robust data-driven model, the authors performed CFD simulations for 200 combinations of randomly generated data points. The study leverages the collected data to develop a data-centric machine learning model capable of accurately predicting flow parameters for specific sorting efficiencies. Remarkably, the developed model achieved a 92% accuracy in predicting the Channel Reynolds Number during testing. However, it is worth noting that the model currently faces challenges in accurately predicting the required number of loops for efficient sorting.

## 1. INTRODUCTION

In the realm of biomedical research, understanding, and analyzing cells have proven crucial for detecting and prognosing various diseases. For instance, anomalous cells, such as Circulating Tumor Cells (CTCs), can indicate cancer progression and metastasis. Additionally, separating circulating fetal cells (CFCs) allows for non-invasive prenatal diagnosis, while the isolation of stem cells enables effective regenerative medicine. Consequently, cell separation techniques have become essential in biotechnology, cancer research, regenerative medicine, and targeted drug delivery [1].

Particle separation techniques can be categorized as active or passive. Active techniques rely on external force fields, while passive techniques utilize channel geometry and hydrodynamic forces. Traditional sorting methods like centrifugation and fluorescent-activated cell sorting have limitations, including high costs, the need for reagents, and potential damage to cells [2]. Inertial microfluidics, however, has emerged as a promising label-free particle sorting technique that overcomes these limitations. It offers faster sample processing, lower costs, and higher precision. Among the microchannel structures used in inertial microfluidics, the serpentine channel stands out as a viable choice due to its small footprint, ease of parallelization, and potential for high throughput. It has been seen form the works of Zhang et al. [3] that serpentine microchannel can be used as an efficient passive particle sorter, in which parameters like channel Reynolds number ($Re_L$), particle diameter ratio ($n$), channel aspect ratio ($AR$) and number of serpentine loops ($N$) dictates the separation efficiency of particles. This happens due to the variation of the forces like inertial lift, dean drag and centrifugal forces acting on the particles [4].

In this paper, the authors produced a numerical model of a symmetric serpentine microchannel particle sorter. The performance of the microchannel with respect to efficiency ($\eta$) and throughput has been investigated subjecting to variation of the characteristic flow parameters by utilizing a machine learning model. It is difficult to obtain the exact combinations of mentioned flow parameters in order to efficiently separate user specific particle combinations, as it requires numerous trial simulations. Thus, a machine learning model is implemented to overcome this computational expense. CFD simulation results corresponding to randomly generated 200 datapoints have been used to train and develop a data-centric ML model, to predict the required channel Reynolds number and the minimum number of loops based on the required sorting efficiency and throughput for a specific diameter ratio.

## 2. PROBLEM FORMULATION AND MODELLING

### 2.1. Geometry and Boundary Conditions

This paper presents an inertial focusing in a two-dimensional microfluidic channel, initially having $AR$ = 0.25 and 30 loops, as shown in Fig. 1. It has a 2D width of 200 $\mu m$, and is considered to have a depth of 50 $\mu m$ in 3D. The outlet is bifurcated for a separate collection of particles. Fig. 1(a) depicts the working of the serpentine microchannel, and zoomed in plot of the inlet and outlet is shown in Fig. 1(b), 1(c). Later the computational domain of the microchannel were altered as the number of loops was varied, keeping other dimensions unaltered, in order to study the effect of variation of number of loops on particle focusing and separation efficiency.

For accurate modelling of the fluid-particle interaction, the particle release time was selected which enabled each particle to be treated as the instantaneous position of one or more particles. All the particles were released into the flow at $t = 1\ sec$, so that the steady flow regime is achieved before the particle release. Along with this, the particle diameter ratio defined as $n = \frac{d_{large}}{d_{small}}$, (ratio of the diameter of the largest and smallest particle), is also crucial in reflecting the particle behavior which in turn affects the flow in the confined microchannel. For every simulation, only 2 types of particles are present at a time with 1000 particles each. In this study a wide range of particle diameter ratio is considered, varying from 1.10 to 2.00, where 7.5 µm is considered as the base diameter. Shown in Fig. 1(b), 1(c), at the microchannel walls, no-slip and particle bounce boundary conditions were applied, and then the calculated flow field was used to trace the particles. At the outlets, zero pressure boundary conditions were imposed.

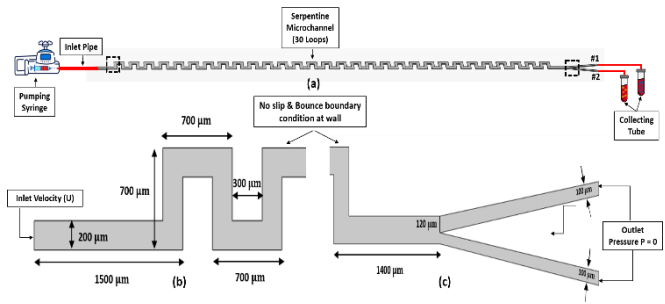

**Figure 1: (a) Schematic of working of a serpentine microchannel particle sorter. (b) Microchannel inlet. (c) Bifurcated microchannel outlet with Boundary conditions.**

### 2.2. Numerical Modelling

In order to model the particles within fluid in the microchannel Discrete Phase Modelling has been utilized. A Eulerian-Lagrangian framework has been used. Due to the small dimensions of a microchannel, relatively low velocity, and consequently low relative Reynolds Number ($Re_L$), the flow field was solved using a laminar flow module in COMSOL [5] by a Eulerian approach. Particles injected at the inlet, which mimic the bio-cells of varying sizes, are solved in the Lagrangian frame. The fluid is considered to be a Newtonian fluid with an average density $\rho = 994\ kg/m^3$ and dynamic viscosity $\mu = 0.0004\ Pa \cdot s$ [6]. The injected particles are treated as a solid particles with fixed shape, size and density of $1110\ kg/m^3$ [7].

An unstructured triangular mesh is used for discretization of the computational domain. A thorough grid-independent and time-independent study has been performed and a mesh size of 109394 elements with a time step of $10^{-4}$ has been identified to be optimum for carrying out the simulation.

Before proceeding with the numerical model for actual study, a validation of the laminar fluid flow and particle tracing model is done against the experimental data of Zhang et al. [3] for inertial focusing in serpentine microchannel. This streak width is plotted against the zigzag loop number, in Fig. 2. The results of the streak width plot obtained from the paper by Zhang et al. [3] are compared with the predicted results of our numerical (CFD) simulation, and the match was found to be good.

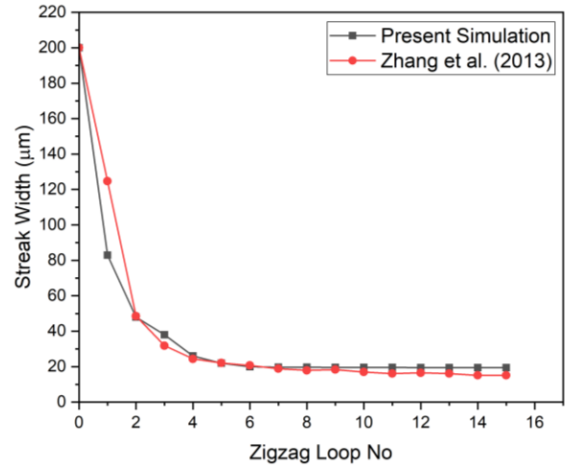

**Figure 2: Streak width vs Loop number plot for particle motion within each loop of the serpentine microchannel, from experimental results of Zhang et al.**

### 2.3 Machine Learning Modelling

Given the high computational load and extensive time consumption involved in the numerical modelling to generate simulation results, in this study we leverage Machine Learning (ML) based techniques to model the relationship between the parameters of the symmetric serpentine microchannel particle sorter. ML approaches are widely recognized in the scientific community for their high accuracy, achieved through their capability to capture and model highly nonlinear behaviours. Unlike traditional models developed from theoretical or established physical reasoning, ML models are "data-driven", built exclusively on observed data. For the particle sorter, we have tried to model the correlation between particle separation efficiency, throughput, flow characteristics (Reynolds number), number of loops and particle diameter ratio through 5 different machine learning based regression techniques. The particle separation efficiency, throughput and particle diameter ratio were considered as the inputs whereas the Reynolds number of the flow and the number of loops were considered as the outputs to the ML models (see Fig. 3). From the developed numerical model (see section 2.2), 200 datapoints were synthesized for the ML techniques. A larger subset (70%) of those data was used for the trainings of the ML algorithms (training data) and the rest data were kept unseen to the trained algorithms and used for the validation of the algorithms (testing data).

To check the stability of the trained models, 50 different random combinations of the testing-training data were tested on the ML models and standard performance metrics were calculated for the assessment of the trained ML models.

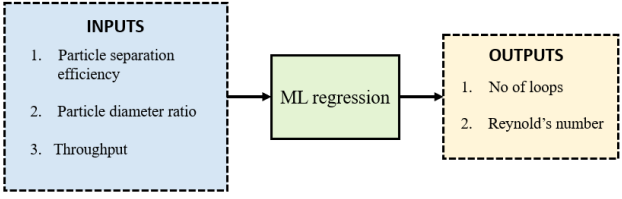

**Figure 3: Machine learning model with input and output features**

**2.4 Machine learning methodology:**

2.4.1 ML models:

In this work, we have considered 3 parameters as inputs and 2 outputs to predict what makes the problem a 'Multi input multi output (MIMO) regression problem. As a benchmark, we started with the Linear regression model, which produced poor accuracy score suggesting a strong nonlinear relationship between the inputs and outputs. Hence, we had utilized 5 other regression techniques; K-nearest neighbors (KNN), Support vector regression (SVR), Random forest regression (RF), Bagging regression (BR), Gradient boosting regression (GBR) that can handle the nonlinear pattern between the multiple inputs and outputs.

2.4.2 Feature scaling:

Due to having different ranges of values in the inputs variables feature scaling technique is used in ML models to standardize the variables, as it results a faster and stable learning. We had incorporated the traditional *StandardScaler* scaling method on the input variables as follow:

$$z_{ji} = \frac{x_{ij} - \mu_i}{\sigma_i}$$

Here, $z_{ji}$ is normalized value of $j^{th}$ observation of feature $i$, $x_{ij}$ is value of $j^{th}$ observation of feature $i$, $\mu_i$ is the mean value of feature $i$, $\sigma_i$ is the standard deviation of feature $i$. First, scaling was done the training data set, then the same scaling parameters were transformed to the testing dataset to prevent snooping/ leakage effects.

2.4.3 Model evaluation metrics:

We used three widely accepted metrics, Mean Absolute Error (MAE), Root Mean Squared Error (RMSE), and Coefficient of Determination ($R^2$ score) to measure the performance of the ML models. Essentially, MAE and RMSE are utilized to capture the difference between the true target values $y_i$ and predicted target values $\hat{y}_i$ over $n$ data points and $R^2$ score is a measure to determine the goodness-of-fit of the regression models; basically, it represents how much variation in the dataset is explained by the regression prediction model.

$$MAE = \frac{1}{n}\sum_{j=1}^{n}|y_i - \hat{y}_i|$$

$$RMSE = \sqrt{\frac{1}{n}\sum_{i=1}^{n}(y_i - \hat{y}_i)^2}$$

$$R^2 score = 1 - \frac{\sum_{i=1}^{n}(y_i - \hat{y}_i)^2}{\sum_{i=1}^{n}(y_i - \bar{y})^2}$$

Overall, MAE, RMSE, $R^2$ score are intuitive metrics that helped us to evaluate the performance of the ML models.

## 3. RESULTS AND DISCUSSION

### 3.1 Separation and Focusing Mechanism:

During the propagation of particles within a serpentine microchannels they mainly experience Drag force, Lift forces, and Centrifugal force [H]. The phenomenon of migration of microparticles within a channel under the influence of these competing forces, into a single focused streak is called focusing. Depending on particle size, they get focused in differential equilibrium positions, and can be separated by appropriate outlet bifurcation. Fig. 4 (a) shows the change in particle separation efficiency with Flow Rate (Q), which in turn depends on Channel Reynolds Number. Fig. 4(b) depicts the variation of separation efficiency with number of loops, for the particle diameter ratio 2, at an optimized flow rate of 250 µL/min.

(a)
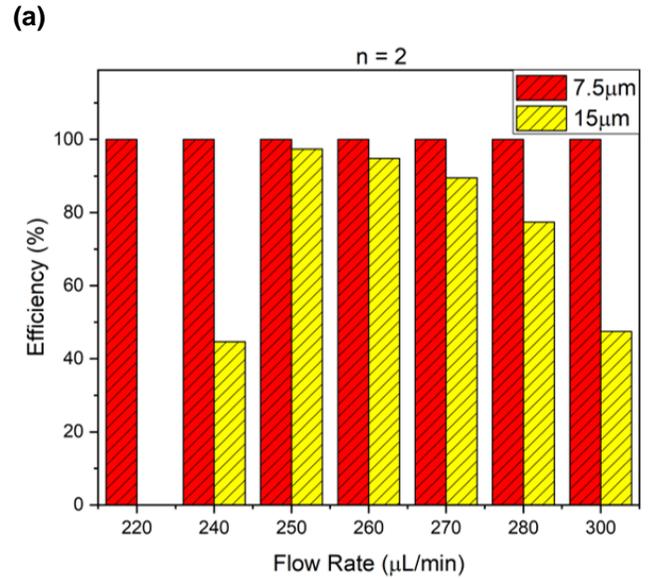

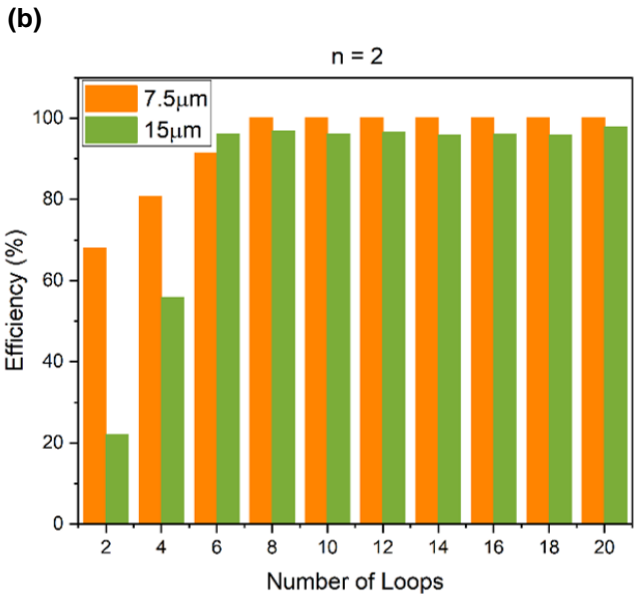

**Figure 4:** (a) Plot for Flow Rate (Q) vs Separation efficiency. (b) Plot for Number of Loops vs Separation efficiency.

It is evident from the Fig. 4(a) that, for a particular diameter ratio of n = 2, the separation efficiency steadily increases with the Flow rate (Q), upto a certain limit. This can be accounted due to the fact that with increase in the flow rate the inertial forces also increases, which in turn increases the lift force on the particles, thus facilitating inertial migration within the microchannel.

Again form Fig. 4(b), an effective steady separation efficiency can only be observed when the number of loops is 6 or more. This can be explained in the context of centrifugal force acting on the particles at each bend of the serpentine channel. Thus, with increase in loops, the particle gets more and more deflected, which leads to focusing and eventually separation [H].

Form the above data it is seen that in case of particular diameter ratio 2, where 7.5 μm is the base diameter, we need an optimized flow rate of 250 μL/min and a minimum of 6 looped serpentine microchannel, for an efficient particle sorting. Thus, it is computationally very tiresome to determine these optimized flow conditions for a particular particle size combination, as it involves a large number of trials.

Hence, the effects of channel Reynolds number ($Re_L$), particle diameter ratio and number of serpentine loops on separation efficiency and throughput, have been analysed using a ML model. The channel Reynolds number is varied from 50-200, Number of Serpentine Loops varied from 2-20 and particle diameter ratio varied from 1.10 to 2.00. The machine learning methodology has been elaborated in section 2.4, and the corresponding results are discussed in the subsequent sections.

## 3.2 Machine Learning Results:

As previously mentioned, we carried out evaluations of 50 different train-test data splits across the five machine learning models. From these 50 splits, mean values of the MAE and RMSE error metrics for the two output variables - Reynolds number and the number of loops - are tabulated in Table 1 & 2. Notably, ensemble learning techniques such as Gradient Boosting Regression (GBR), Bagging Regression (BR), and Random Forest (RF) demonstrated superlative performance in comparison to the K-Nearest Neighbors (KNN) and Support Vector Regression (SVR) methods. This superior performance can be attributed to the dataset size; when the dataset is relatively small, conventional regression methods tend to overfit, thus generating higher errors on the test dataset. On the contrary, ensemble learning approaches have the potential to enhance model generalization and mitigate overfitting. They achieve this by creating a variety of training subsets, ensuring the model learns from assorted perspectives, and avoiding overfitting to the original dataset.

**Table 1: performance error metrics on Re target variable**

| ML model | MAE | | RMSE | |
|---|---|---|---|---|
| | Train data | Test data | Train data | Test data |
| KNN | 12.96 | 16.57 | 17.49 | 21.85 |
| RF | 3.63 | 9.40 | 5.10 | 12.94 |
| BR | 5.97 | 9.97 | 8.11 | 13.33 |
| GBR | 6.12 | 10.03 | 7.44 | 12.80 |
| SVR | 29.45 | 29.80 | 34.97 | 35.21 |

**Table 2: performance error metrics on No of loops target variable**

| ML model | MAE | | RMSE | |
|---|---|---|---|---|
| | Train data | Test data | Train data | Test data |
| KNN | 2.46 | 3.09 | 3.21 | 3.93 |
| RF | 1.19 | 3.00 | 1.60 | 3.77 |
| BR | 1.93 | 3.17 | 2.37 | 3.84 |
| GBR | 1.56 | 2.79 | 1.98 | 3.60 |
| SVR | 2.95 | 3.35 | 3.69 | 4.01 |

Additionally, we evaluated the coefficient of determination ($R^2$ score) for both the training and testing datasets corresponding to the two output variables. The Fig. 6 clearly indicate that the machine learning models are more adept at fitting the Reynold's number ($Re_L$) than the 'Number of loops' output variable. The $Re_L$ output exhibits less variation (i.e., more stable) in the $R^2$ score values compared to the 'Number of loops', which can be ascribed to the continuous nature of the $Re_L$ compared to the discrete integer datatype of the 'Number of loops' variable. Among all the techniques, the Gradient Boosting Regression (GBR) outperforms others with a mean $R^2$ score of 0.98 (training) and 0.92 (testing) for $Re_L$, and 0.86 (training) and 0.51 (testing) for the 'Number of loops'. The Fig. 5 demonstrates the actual versus predicted plot for the output variables using GBR, reinforcing that the trained model tends to correspond closely with the best fit line.

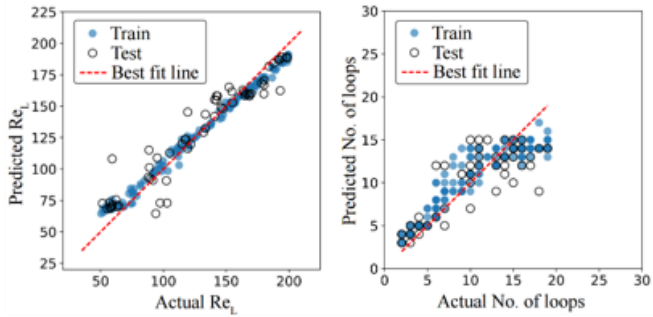

**Figure 5: Actual vs predicted Reynolds number and Number of Loops, using GBR.**

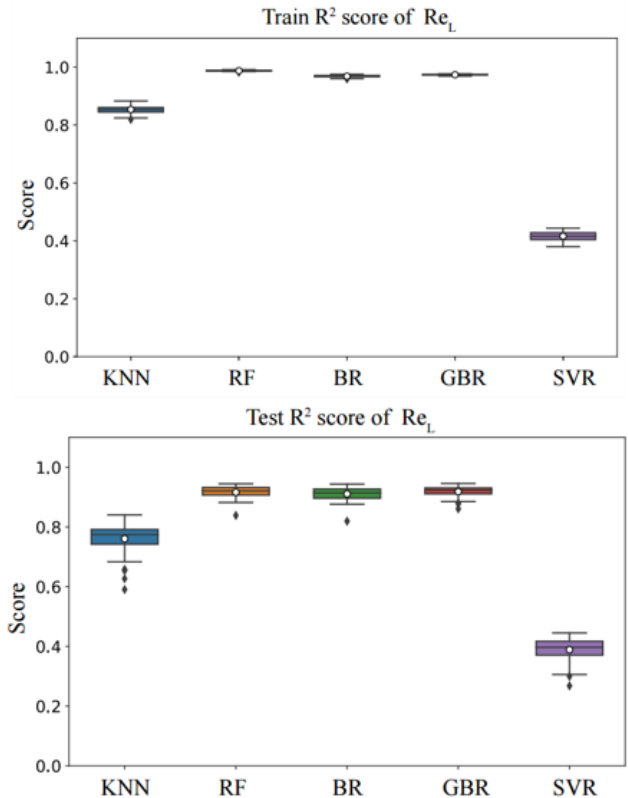

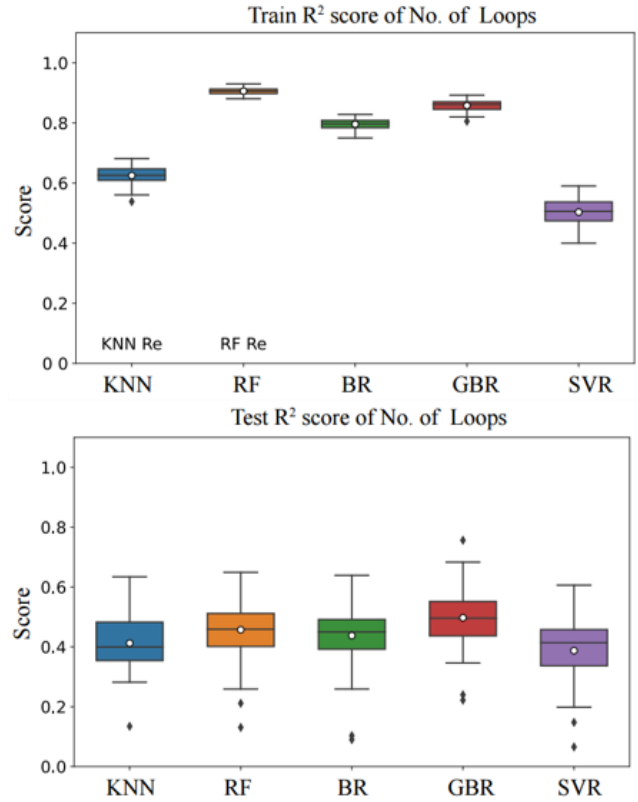

**Figure 6**: $R^2$ score comparison for Reynolds number and Number of Loops.

## 4. CONCLUSION

In this research endeavor, we investigated the sorting efficiency and throughput of a symmetric serpentine microchannel particle sorter subjecting the channel to various flow parameters and henceforth applied a data-driven machine learning approach to model the correlation between the flow parameters and both efficiency and throughput. Our comparative analysis of 5 distinct non-parametric machine learning regression techniques revealed ensemble-based regression methods as the most effective in predicting the ideal Reynolds number and the number of loops for achieving a targeted particle separation efficiency. In particular, using the Gradient Boosting Regression method, we achieved a mean $R^2$ score accuracy of 98% in training and 92% in testing for predicting Reynolds number, and a mean $R^2$ score accuracy of 86% in training and 51% in testing for predicting the number of loops. The relatively lower prediction accuracy for the number of loops, attributable to its discrete datatype, will be addressed with more suitable machine learning methodologies in future stages of this work.

### NOMENCLATURE

| | | |
|---|---|---|
| $Re_L$ | Channel Reynolds Number | -- |
| AR | Aspect ratio | -- |
| n | Number of Loops | -- |
| η | Particle Separation Efficiency | -- |

| Symbol | Description | Units |
|---|---|---|
| $\rho, \rho_f$ | Liquid density | [kg/m$^3$] |
| $\rho_p$ | Particle density | [kg/m$^3$] |
| $\mu$ | Dynamic viscosity of the fluid | [kg/m.s] |
| $U_m$ | Channel inlet velocity | [m/s] |
| $a$ | Particle diameter | [m] |
| $D_h$ | Hydraulic channel diameter | [m] |
| $R_p$ | Particle Reynolds Number | -- |
| $f_c$ | Lift Coefficient | -- |
| $r$ | Radius of curvature | [m] |
| $v_{fr}$ | Radial fluid velocity | [m/s] |
| $v_{pr}$ | Radial particle velocity | [m/s] |
| $v_{pt}$ | Tangential particle velocity | [m/s] |
| $F_L$ | Inertial Lift Force | [N] |
| $F_D$ | Drag Force | [N] |
| $F_{Cent}$ | Centrifugal Force | [N] |